\documentclass[twocolumn,english,prb,showpacs,amsmath,amssymb,eqsecnum,preprintnumbers]{revtex4}
\usepackage[T1]{fontenc}
\usepackage[latin9]{inputenc}
\usepackage{color}
\usepackage{bm}
\usepackage{amstext}
\usepackage{esint}
\usepackage{soul}% \st{text} provides strikeout of "text"
\usepackage{graphicx}
\usepackage{dcolumn}
\usepackage{bigstrut}
\usepackage{rotating}    

\makeatletter
\@ifundefined{textcolor}{}
{%
 \definecolor{BLACK}{gray}{0}
 \definecolor{WHITE}{gray}{1}
 \definecolor{RED}{rgb}{1,0,0}
 \definecolor{GREEN}{rgb}{0,1,0}
 \definecolor{BLUE}{rgb}{0,0,1}
 \definecolor{CYAN}{cmyk}{1,0,0,0}
 \definecolor{blue}{cmyk}{0,1,0,0}
 \definecolor{YELLOW}{cmyk}{0,0,1,0}
}

\def\chiL{\chi_{\text{L}}}

\def\sgn{{\text{sgn\,}}}

\def\be{\begin{equation}}
\def\ee{\end{equation}}
\def\bea{\begin{eqnarray}}
\def\eea{\end{eqnarray}}
\def\bse{\begin{subequations}}
\def\ese{\end{subequations}}

\def\bml{\begin{mathletters}}
\def\eml{\end{mathletters}}
\def\NLSM{NL$\sigma$M}
\usepackage{dcolumn}% Align table columns on decimal point
\usepackage{bm}% bold math
\usepackage{amsfonts}
\usepackage[normalem]{ulem}
\draft

\makeatother

\usepackage{babel}
\begin{document}
%\preprint{Phys. Rev. B {\bf 91}, 214407 (2015)}

\title{Magnon-induced long-range correlations and their neutron-scattering signature in quantum magnets}

\author{S. Bharadwaj$^1$, D. Belitz$^{1,2}$, and T. R. Kirkpatrick$^{3}$}

\affiliation{ $^{1}$ Department of Physics and Institute of Theoretical Science, University of Oregon, Eugene, OR 97403, USA\\
                  $^{2}$ Materials Science Institute, University of Oregon, Eugene, OR 97403, USA\\
                  $^{3}$ Institute for Physical Science and Technology, University of Maryland, College Park, MD 20742, USA\\
}

\date{\today}
\begin{abstract}
We consider the coupling of the magnetic Goldstone modes, or magnons, in both quantum ferromagnets and antiferromagnets to
the longitudinal order-parameter fluctuations, and the resulting nonanalytic behavior of the longitudinal susceptibility. In classical
magnets it is well known that long-range correlations induced by the magnons lead to a singular wave-number dependence of
the form $1/k^{4-d}$ in all dimensions $2<d<4$, for both ferromagnets and antiferromagnets. At zero temperature we find a 
profound difference between the two cases. Consistent with naive power counting, the longitudinal susceptibility in a quantum
antiferromagnet scales as $k^{d-3}$ for $1<d<3$, whereas in a quantum ferromagnet the analogous result, $k^{d-2}$, is absent 
due to a zero scaling function. This absence of a nonanalyticity in the longitudinal susceptibility is due to the lack of magnon 
number fluctuations in the ground state of a quantum ferromagnet; correlation functions that are sensitive to other fluctuations 
do exhibit the behavior predicted by simple power counting.  Also of interest is the dynamical behavior as expressed in the 
longitudinal part of the dynamical structure factor, which is
directly measurable via neutron scattering. For both ferromagnets and antiferromagnets there is a logarithmic singularity
at the magnon frequency with a prefactor that vanishes as $T\to 0$. In addition, in the antiferromagnetic case there is a
nonzero contribution at $T=0$ that is missing for ferromagnets. Magnon damping due to quenched disorder restores the expected
scaling behavior of the longitudinal susceptibility in the ferromagnetic case; it scales as $k^{d-2}$ if the order parameter is not 
conserved (magnetic disorder), or as $k^d$ if it is (non-magnetic disorder). Detailed predictions are made for both two- and 
three-dimensional systems at both $T=0$ and in the limit of low temperatures, and the physics behind the various nonanalytic 
behaviors is discussed.
\end{abstract}
\maketitle

\section{Introduction}
\label{sec:I}

The collective excitations known as magnons are a characteristic feature of any magnetically ordered state in which
a continuous symmetry is spontaneously broken.\cite{Chaikin_Lubensky_1995} Common examples are planar, or XY, and Heisenberg magnets,
where the spontaneously broken symmetry is $O(2)$ and $O(3)$, respectively. The magnons are the resulting
Goldstone modes, which are soft or massless since a uniform rotation of the order parameter does not require
any energy. In a ferromagnet, their frequency $\Omega$ scales as the wave number $k$ squared in the long-wavelength limit,
$\Omega \sim k^2$; in an antiferromagnet, the frequency is a linear function of the wave number, $\Omega \sim k$.
The relevant correlation function is the transverse order-parameter susceptibility, which diverges in the limit of
zero frequency and wave number. In a solid, the magnons are gapped at asymptotically small frequencies, and the 
transverse susceptibility stays finite, due to the underlying lattice that breaks
the $O(n)$ symmetry; however, compared to other relevant energy scales this is usually a small effect due to the
weakness of the spin-orbit interaction. For our purposes we will ignore the spin-orbit interaction %unless otherwise noted 
and treat the magnons as gapless. 

\begin{figure}[t]
\includegraphics[width=4cm]{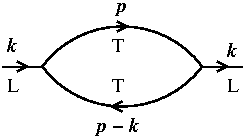}
\caption{Diagrammatic representation of the coupling between longitudinal and transverse spin flucutuations in 
              the classical case: A longitudinal (L) mode couples
              to two transverse (T) modes. The resulting contribution to the longitudinal
              susceptibility $\chi_{\text{L}}$ has the form given in Eq.~(\ref{eq:1.1}).}
\label{fig:1}
\end{figure}

Magnons can be observed directly via neutron scattering.\cite{Forster_1975, Chaikin_Lubensky_1995} However, 
via couplings of the transverse order-parameter
fluctuations to other modes, they also have profound indirect effects on other observables. An example is the 
longitudinal spin susceptibility $\chiL$ in a classical Heisenberg ferromagnet or antiferromagnet. It has been known for a long time that
the coupling of the longitudinal spin fluctuations to the transverse ones (i.e., the Goldstone modes) leads to a $\chiL$ that diverges
for $k\to 0$ everywhere in the ordered phase for all spatial dimensions $2<d<4$.\cite{Vaks_Larkin_Pikin_1967,
Brezin_Wallace_1973, 2d_footnote} The leading contribution takes the form of a convolution of two Goldstone modes
\be
\chiL \propto \int d{\bm p}\,\frac{1}{{\bm p}^2}\,\frac{1}{({\bm p}-{\bm k})^2} \approx \int_k d^dp\,\frac{1}{p^4} \propto \frac{1}{k^{4-d}}\ .
\label{eq:1.1}
\ee
It can be represented diagrammatically as shown in Fig.~\ref{fig:1}.
This result, which was originally derived for ferromagnets in perturbation theory, was later shown by renormalization-group (RG) methods
to be asymptotically exact.\cite{Belitz_Kirkpatrick_1997} It reflects the scale dimensions that characterize the stable RG
fixed point that describes the ordered phase. We stress again that Eq.~(\ref{eq:1.1}) holds for
both classical ferromagnets and antiferromagnets. However, in the latter the physical
meaning of the longitudinal order-parameter susceptibility $\chiL$ is the correlation function of the
staggered magnetization, rather than the spin susceptibility.

Physically, the nonanalytic dependence of $\chiL$ on the wave number
reflects long-range correlations in the system that are due to the massless magnons: In real space, $\chiL$ for
large distances $r$ falls off as a power law, $\chiL \propto 1/r^{2d-4}$. This is a particular manifestation of a
more general phenomenon: Soft or massless modes lead to long-range correlations that are reflected in
nonanalytic wave-number and frequency dependences in the hydrodynamic limit, i.e., the limit
of small frequencies and wave numbers, in observables that couple to the soft modes.
If the soft modes exists in entire phases, as opposed to, e.g., isolated critical points, then so does the
nonanalytic behavior, which usually takes the form of power laws; a phenomenon known as generic scale 
invariance.\cite{Belitz_Kirkpatrick_Vojta_2005} In magnets, other soft modes may be present that also couple 
to any given observable and compete with the magnons in producing long-range correlations, or nonanalytic 
behavior. For instance, in disordered metals at zero temperature ($T=0$) there are 
diffusive excitations known as ``diffusons'' and ``Cooperons'' that lead to nonanalyticities in observables known
as weak-localization effects.\cite{Altshuler_Aronov_1984, Lee_Ramakrishnan_1985}  A specific example is the
nonanalytic frequency dependence of the electrical conductivity. The competing effects of
magnons on one hand, and of diffusons and Cooperons on the other, on the conductivity in disordered
metallic ferromagnets have been investigated in Ref.~\onlinecite{Kirkpatrick_Belitz_2000}.

The weak-localization and other zero-temperature effects raise an interesting question: What is the fate
of the singular behavior of the classical longitudinal spin susceptibility, Eq.~(\ref{eq:1.1}), in the limit $T\to 0$?
Simple considerations show that the singularity must be weaker at $T=0$. In quantum statistical mechanics
the statics and the dynamics are intrinsically coupled. The expression for $\chiL$ at $T=0$ therefore must 
include a frequency integration in addition to the wave-number integration, and the integrand must be comprised
of the dynamic Goldstone modes. A simple guess, based on power counting only, is that for quantum
antiferromagnets at $T=0\ $~\cite{notation_footnote}
\bse
\label{eqs:1.2}
\be
\chiL \sim \int_k d^dp \int_{\Omega} d\omega\ \frac{1}{(p^2+\omega^2)^2} \sim k^{d-3} \sim \Omega^{d-3}
\label{eq:1.2a}
\ee
for $1<d<3$, with a logarithmic singularity in $d=3$.
We will show below that this expectation is indeed borne out by an explicit calculation. 

For quantum ferromagnets, the corresponding expression obtained by replacing the denominator in Eq.~(\ref{eq:1.2a}) 
by $(p^2 + \omega)^2$ is clearly not correct. This can be seen from spin-wave theory, which expresses
the spin operators by bosonic operators via a Holstein-Primakoff transformation.\cite{Kittel_1963}
In a ferromagnet, the longitudinal spin is given in terms of the magnon-number operator, and $\chi_{\text{L}}$
thus is the magnon-number correlation function. At $T=0$ there are no magnons, and the contribution
corresponding to Eq.~(\ref{eq:1.2a}) (which would scale as $k^{d-2}$) therefore has a zero prefactor. 
An equivalent statement is that in the ground state of a quantum ferromagnet the magnetization has
its maximum value, and therefore the ground-state energy has the same value as it does classically and
cannot be decreased by quantum fluctuations.\cite{FM_caveat} In a quantum antiferromagnet, by contrast, this is not true: 
The classical Ne{\'e}l state is not an eigenvalue of the Hamiltonian, and the ground-state energy is lowered 
below its classical value by quantum fluctuations. The remaining question is how the classical singularity, Eq.~(\ref{eq:1.1}), 
vanishes as $T\to 0$ in a ferromagnet. As we will see, the leading contribution for $k\to 0$ at a low fixed temperature 
is given by  Eq.~(\ref{eq:1.1}) with a $T$ prefactor,
\be
\chiL \propto T \int d{\bm p}\,\frac{1}{{\bm p}^2}\,\frac{1}{({\bm p}-{\bm k})^2} \propto \frac{T}{k^{4-d}}\ .
\label{eq:1.2b}
\ee
\ese

The above considerations hold for undamped magnons. If the magnons are damped, then in general
a nonanalyticity in the hydrodynamic limit is restored, with the exponent depending on the nature of
the damping.\cite{damping_footnote} For instance, magnetic impurities, which lead to a damping
coefficient proportional to $p^2$, introduce sufficiently strong fluctuations to invalidate the arguments
given below Eq.~(\ref{eq:1.2a}) and lead to a longitudinal susceptibility that does indeed scale as
$k^{d-2}$ at $T=0$. Non-magnetic quenched disorder, which leads to a damping coefficient proportional
to $p^4$, leads to a weaker singularily, $\chi_{\text{L}} \sim k^d$.

A more general question pertains to the spectrum of the dynamical longitudinal susceptibility or, equivalently,
the longitudinal part of the dynamical structure factor, which is directly measurable by neutron scattering,
as is the transverse part. For bulk ferromagnets at $T>0$,
the longitudinal structure factor has a logarithmic singularity at the magnon resonance, which gets regularized by a
magnetic field.\cite{Vaks_Larkin_Pikin_1967} We will show that for an antiferromagnet there is a nonzero
contribution even at $T=0$, which is caused by the same quantum fluctuations that are responsible for
Eq.~(\ref{eq:1.2a}) to hold. The singularity at the magnon resonance takes the form of a discontinuous slope
in bulk antiferromagnets, and a square-root singularity in two-dimensional systems.

The organization of the paper is as follows. In Sec.~{\ref{sec:II} we consider magnets with undamped spin waves by considering
nonlinear sigma models (\NLSM s) for both quantum ferromagnets and antiferromagnets. This provides a simple and transparent way to
understand why the classical nonanalyticity disappears as $T\to 0$ in the ferromagnetic case, while it is just weakened, in
agreement with the simple scaling argument given above, in the antiferromagnetic case. In Sec.~\ref{sec:III} we use
time-dependent Ginzburg-Landau theory to discuss the effects of damped magnons in ferromagnets. In Sec.~\ref{sec:IV} 
we conclude with a summary and discussion of our results.

\section{Effects of undamped magnons}
\label{sec:II}

Nonlinear sigma models (\NLSM s) provide a convenient description of the long-wavelength and low-frequency properties 
of the ordered phase of systems with a 
spontaneously broken symmetry. They are effective field theories that focus on the Goldstone modes and integrate out all massive 
fluctuations in the simplest approximation that respects the symmetry. In particular, the classical $O(3)$-symmetric nonlinear sigma 
model \cite{Zinn-Justin_1996} provides a very easy way to demonstrate the divergence of $\chi_{\text{L}}$ in a classical Heisenberg 
ferromagnet, Eq.~(\ref{eq:1.1}). It thus is natural to consider quantum \NLSM s to study the corresponding effect in quantum magnets. 
As we will see, the results are very different for the two types of magnetic order.

\subsection{Quantum ferromagnets}
\label{subsec:II.A}

We consider a quantum ferromagnet with a fluctuating magnetization ${\bm M}(x) = M_0(x)\,{\hat{\bm m}}(x)$. Here and in
what follows $x = ({\bm x},\tau)$ comprises the real-space position ${\bm x}$ and the imaginary-time variable $\tau$.
$M_0$ is the magnitude of the order parameter, and
\bse
\label{eqs:2.1}
\be
{\hat{\bm m}}(x) = \left(\pi_1(x),\pi_2(x),\sigma(x)\right)
\label{eq:2.1a}
\ee
with
\be
{\hat{\bm m}}^2(x) = \pi_1^2(x) + \pi_2^2(x) + \sigma^2(x) \equiv 1
\label{eq:2.1b}
\ee
\ese
is a unit vector. In a \NLSM\ description of a quantum ferromagnet fluctuations of $M_0$ are neglected,
$M_0(x) \equiv M_0$, and the partition function can be written \cite{Fradkin_1991, Sachdev_1999}
\bse
\label{eqs:2.2}
\be
Z = \int {\cal D}[\hat{\bm m}]\ \delta(\hat{\bm m}^2(x) - 1)\ e^{-\int dx\,{\cal L}_{\text{FM}}[\hat{\bm m}]}
\label{eq:2.2a}
\ee
Here $\int dx = \int_0^{1/T} d\tau \int_V d{\bm x}$, with $T$ the temperature and $V$ the system volume, and
\bea
{\cal L}_{\text{FM}}[\hat{\bm m}] &=&  -\frac{\rho_s}{2}\, \hat{\bm m}(x)\cdot{\bm\nabla}^2\hat{\bm m}(x) - M_0\,\mu{\bm H}\cdot\hat{\bm m}(x)
\nonumber\\
&& \hskip -25pt + \frac{i M_0}{1+\sigma(x)} \bigl(\pi_1(x)\partial_{\tau} \pi_2(x) - \pi_2(x)\partial_{\tau} \pi_1(x)\bigr)\ .
\nonumber\\
\label{eq:2.2b}
\eea
\ese
Here $\rho_s$ is the spin-stiffness coefficient, which is proportional to $M_0^2$, $\bm H$ is an external magnetic field,
and $\mu$ is the gyromagnetic ratio. 
The first two terms on the right-hand side of Eq.~(\ref{eq:2.2b}) are the same as in a classical $O(3)$ \NLSM.\cite{Zinn-Justin_1996} 
The third term is the Wess-Zumino or Berry-phase term that describes the quantum dynamics.\cite{WZ_footnote} Physically, it 
describes the Bloch spin precession. The form given in Eq.~(\ref{eq:2.2b}) assumes that the ferromagnet order is along the $z$-direction.

We now expand the action in powers of the fields $\pi_1$ and $\pi_2$. The Gaussian action that governs the transverse fluctuations 
then reads
\bse
\label{eqs:2.3}
\be
{\cal A}^{(2)}[\pi_1,\pi_2] = \frac{M_0}{2} \sum_k \sum_{i,j=1}^{2} \pi_i(k)\,\Gamma_{ij}(k)\,\pi_j(-k)\ ,
\label{eq:2.3a}
\ee
where $\Gamma_{ij}$ denotes the matrix elements of a $2\times 2$ matrix
\be
\Gamma(k) = \begin{pmatrix} D\, {\bm k}^2 + \mu H  &  -\Omega_n   \\
                                                     \Omega_n    &  D\, {\bm k}^2 + \mu H      \end{pmatrix}
\label{eq:2.3b}
\ee
\ese 
where $D = \rho_s/M_0$. Here we have performed a Fourier transform from $x = ({\bm x},\tau)$ to $k = ({\bm k},i\Omega_n)$, 
with ${\bm k}$ a wave vector and $\Omega_n = 2\pi T n$ ($n$ integer) a bosonic Matsubara frequency, and we
have taken the external field to point in the $z$-direction, ${\bm H} = (0,0,H)$. The inverse of $\Gamma$ yields the 
Gaussian transverse susceptibility matrix, i.e., the correlation function
\bse
\label{eqs:2.4}
\be
M_0^2 \langle\pi_i(k)\,\pi_j(-k)\rangle = \chi_{\text{T}}^{ij}(k)\ ,
\label{eq:2.4a}
\ee
where
\be
\chi_{\text{T}}(k) = \frac{M_0}{(D{\bm k}^2 + \mu H)^2 + \Omega_n^2}
                             \begin{pmatrix}D{\bm k}^2 + \mu H  &  \Omega_n   \\
                                                    -\Omega_n    &  D{\bm k}^2  + \mu H      \end{pmatrix}
\label{eq:2.4b}
\ee
\ese
The non-hermitian nature of the matrix $\Gamma$, with the frequency coupling the magnetization components 
$M_x$ and $M_y$, reflects the structure of the Bloch spin-precession term in Eq.~(\ref{eq:2.2b}). It shows the
quadratic dispersion relation of the ferromagnetic magnons, $i\Omega_n = \pm D{\bm k}^2$. The spin-wave
stiffness coefficient $D$ (not to be confused with a diffusion coefficient) is linear in $M_0$ (since 
$\rho_s \propto M_0^2$). It is illustrative to diagonalize the Gaussian transverse action. The eigenvalues of 
$\Gamma(k)$ are $\lambda_{\pm}(k)$ with
\bse
\label{eqs:2.5}
\be
\lambda_{\pm}(k) = \lambda_{\mp}(-k) = D{\bm k}^2 + \mu H \mp i\Omega_n\ ,
\label{eq:2.5a}
\ee
and the left and right eigenvectors are
\bea
(u,v)_L &=& (1,\mp i)\ ,
\nonumber\\
(u,v)_R &=& (1,\pm i)\ .
\label{eq:2.5b}
\eea
\ese
The Gaussian action can thus be written in terms of fields $\psi_L = (\psi_{L,+},\psi_{L,-})$ and $\psi_R = (\psi_{R,+},\psi_{R,-})$,
\be
{\cal A}^{(2)}[\psi_L,\psi_R] = \frac{M_0}{2}\sum_k\sum_{\sigma = \pm} \psi_{\text{L},\sigma}(k)\,\lambda_{\sigma}(k)\,\psi_{\text{R},\sigma}(-k)
\label{eq:2.6}
\ee    
In terms of the $\psi_L$ and $\psi_R$ we have
\bea
\pi_1 &=& \frac{1}{\sqrt{2}}\left(\psi_{L,+} - i\psi_{L,-}\right) = \frac{1}{\sqrt{2}}\left(\psi_{R,+} + i\psi_{R,-}\right)\ ,
\nonumber\\
\pi_2 &=& \frac{1}{\sqrt{2}}\left(-i\psi_{L,+} + \psi_{L,-}\right) = \frac{1}{\sqrt{2}}\left(i\psi_{R,+} + \psi_{R,-}\right)\ .
\nonumber\\
\label{eq:2.7}
\eea    
Note that the four fields $\psi_{\text{L},\sigma}$, $\psi_{\text{R},\sigma}$
are not independent; Eqs.~(\ref{eq:2.7}) yield two constraints,
\be
\psi_{\text{L},+} = i\psi_{\text{R},-}\quad,\quad \psi_{\text{L},-} = i\psi_{\text{R},+}\ ,
\label{eq:2.8}
\ee
which restore the original number of degrees of freedom.
From Eq.~(\ref{eq:2.6}) we obtain the Goldstone mode\cite{FM_GM_footnote}
\bse
\label{eqs:2.9}
\be
g_{\pm}(k) = \langle\psi_{L,\pm}(k) \psi_{R,\pm}(-k)\rangle = 1/M_0\,\lambda_{\pm}(k)
\label{eq:2.9a}
\ee
which is massless in the absence of the symmetry-breaking field $H$. From Eq.~(\ref{eq:2.8}) we obtain two
additional nonzero correlation functions,
\bea
\langle\psi_{\text{L}+}(k)\,\psi_{\text{L},-}(-k)\rangle &=& i/M_0\,\lambda_{+}(k)\ ,
\nonumber\\
\langle\psi_{\text{R}+}(k)\,\psi_{\text{R},-}(-k)\rangle &=& -i/M_0\,\lambda_{+}(-k)\ ,
\label{eq:2.9b}
\eea
\ese

Now we consider the normalized longitudinal susceptibility $\chi_{\text{L}}(x-y)/M_0^2 = \langle\delta\sigma(x)\delta\sigma(y)\rangle$
with $\delta\sigma(x) = \sigma(x) - \langle\sigma(x)\rangle$. Using the
nonlinear constraint, Eq.~(\ref{eq:2.1b}), we expand
\begin{widetext}
\bse
\label{eqs:2.10}
\be
\langle \delta\sigma(x) \delta\sigma(y) \rangle = \frac{1}{4} \langle (\pi_1^2(x) + \pi_2^2(x))( \pi_1^2(y) + \pi_2^2(y)) \rangle 
 -  \frac{1}{4} \langle \pi_1^2(x) + \pi_2^2(x)\rangle^2 + \ldots
\label{eq:2.10a}
\ee
In terms of $\psi_L$ and $\psi_R$ this can be written
\be
\langle \delta\sigma(x) \delta\sigma(y) \rangle = \frac{1}{4} \sum_{\sigma,\sigma'} \Bigl[ \langle \psi_{L,\sigma}(x)\,\psi_{R,\sigma}(x)\,
     \psi_{L,\sigma'}(y)\,\psi_{R,\sigma'}(y)\rangle
 -  \langle \psi_{L,\sigma}(x)\,\psi_{R,\sigma}(x)\rangle \langle \psi_{L,\sigma'}(y)\,\psi_{R,\sigma'}(y)\rangle   \Bigr]\ .
\label{eq:2.10b}
\ee
\ese
\end{widetext}
Using Wick's theorem and Eqs.~(\ref{eqs:2.9}) yields the one-loop contribution $\chi_{\text{L}}^{(1)}$ to the longitudinal
susceptibility,
\bea
\chi_{\text{L}}^{(1)}(k) &=& M_0^2\,\frac{T}{2V}\sum_p \sum_{\sigma} g_{\sigma}(p)\,g_{\sigma}(p-k)
\nonumber\\
&=& \frac{T}{2V}\sum_p \sum_{\sigma} \frac{1}{\lambda_{\sigma}(p) \lambda_{\sigma}(p-k)}\ .  \qquad                                        
\label{eq:2.11}
\eea
This is represented diagrammatically in Fig.~\ref{fig:2}.

\subsubsection{Absence of a Goldstone-mode-induced singularity in $\chi_{\text{L}}$ at $T=0$}
\label{subsubsec:II.A.1}

\begin{figure}[b]
\includegraphics[width=4cm]{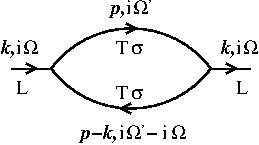}
\caption{Diagrammatic representation of the coupling between longitudinal and transverse spin flucutations
              in the quantum case. Note that the two transverse
              propagators carry the same internal frequency. This leads to the null result discussed in the text.}
\label{fig:2}
\end{figure}
At $T=0$, where the frequency summation in Eq.~(\ref{eq:2.11}) turns into an integral, it is obvious that this contribution
vanishes,\cite{integral_footnote} in violation of the naive expectation expressed by the ferromagnetic analog of Eq.~(\ref{eq:1.2a}). 
This null result is readily traced back to the structure of the Bloch spin precession term in the action, which leads to the eigenvalues
$\lambda_{\sigma}(k)$ being odd functions of the frequency. Since the action couples only $\psi_{L,+}$ with $\psi_{R,+}$,
and $\psi_{L,-}$ with $\psi_{R,-}$, this results in a final frequency integral where both poles lie on the same side of the real axis.
Alternatively, one can easily see this in an operator formalism, see the discussion after Eq.~(\ref{eq:1.2a}) in the Introduction.
Adding a frequency dependence to the classical expression therefore has a much stronger effect
than increasing the effective dimensionality by two, as the naive power-counting argument suggests, and at $T=0$
it completely suppresses the effect. It is obvious from this discussion that the absence of a nonanalyticity
in the quantum case is a generic property of ferromagnets at $T=0$ and not an artifact of either the \NLSM\ or
the one-loop approximation. We also note that the null result is specific to the 2-point correlation of $\sigma(x)$,
see the following section.

\begin{figure*}[h,t]
\includegraphics[width=14cm]{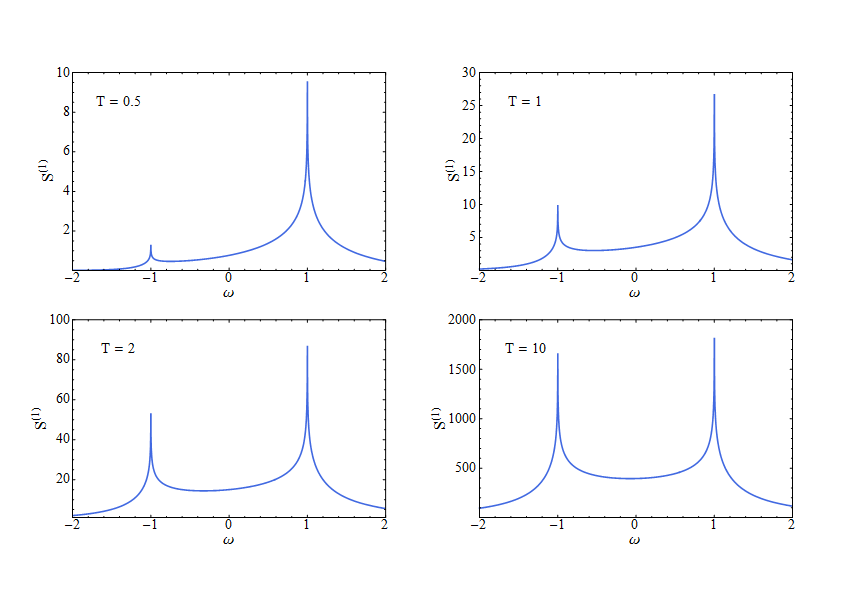}
\caption{The one-loop contribution to the longitudinal part of the dynamical structure factor for a ferromagnet, Eq.~(\ref{eq:2.21}), 
              normalized by 
              $\sqrt{\omega_{\bm k}}/4\pi D^{3/2}$, for $H=0$ as a function of the frequency $\omega$ for various values of the
              temperature $T$. $\omega$ and $T$ are measured in units of $\omega_{\bm k}$. On the scale shown, the 
              result for $T/\omega_{\bm k} = 10$ is almost indistinguishable from the classical result, Eq.~(\ref{eq:2.22}).}
\label{fig:3}
\end{figure*}
\begin{figure*}[h,t]
\includegraphics[width=14cm]{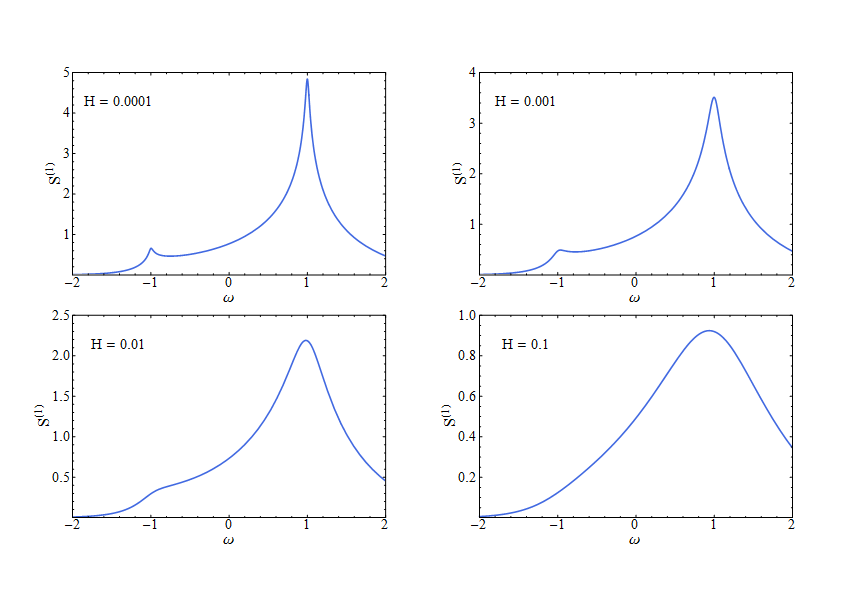}
\caption{The one-loop contribution to the longitudinal part of the dynamical structure factor for a ferromagnet, Eq.~(\ref{eq:2.21}), normalized as in 
              Fig.~\ref{fig:3}, for $T/\omega_{\bm k} = 0.5$ as a function of the frequency $\omega$ for various values of the
              magnetic field $H$. $\omega$ and $H$ are measured in units of $\omega_{\bm k}$ and $\omega_{\bm k}/\mu$,
              respectively. Notice how even a very weak  magnetic field broadens the resonance feature.}  
\label{fig:4}
\end{figure*}

\subsubsection{A singular correlation function at $T=0$}
\label{subsubsec:II.A.1'}

It is illustrative to discuss a correlation function other than $\chi_{\text{L}}$. Consider, for instance,
\bea
\Psi(x-y) &=& \frac{1}{4}\langle (\pi_1^2(x) - \pi_2^2(x))( \pi_1^2(y) - \pi_2^2(y)) \rangle
\nonumber\\
     &&\hskip -45pt =  \frac{1}{4}\sum_{\sigma,\sigma'} \sigma\sigma' \langle \psi_{L,\sigma}(x)\,\psi_{L,\sigma}(x)\,\psi_{R,\sigma'}(y)\,\psi_{R,\sigma'}(y)\rangle.
\nonumber\\
\label{eq:2.12}
\eea
Note that this is a physical, if hard to measure, correlation function: It describes the response to a ``field'' $\Delta$ that renders the
exchange coupling $J$ in a Heisenberg model anisotropic in the $x$-$y$-plane: $J_x = J + \Delta$, $J_y = J - \Delta$. 
After a Fourier transform we obtain, instead of Eq.~(\ref{eq:2.11}),
\be
\Psi(k) =  \frac{T}{2V}\sum_p \sum_{\sigma} \frac{1}{\lambda_{\sigma}(p) \lambda_{\sigma}(k-p)}\ .
\label{eq:2.13}
\ee
At $T=0$, the frequency integral is now over a function that has poles on either side of the real axis, and the correlation
function behaves as simple power counting would suggest, viz.
\bea
\Psi({\bm k},i\Omega_n=0) &\propto& \text{const.} + \vert{\bm k}\vert^{d-2}\ ,
\nonumber\\
\Psi({\bm k}=0,i\Omega_n) &\propto& \text{const.} + \vert\Omega_n\vert^{(d-2)/2}\ ,
\label{eq:2.14}
\eea
with a logarithmic singularity in $d=2$. This is
in complete analogy to Eq.~(\ref{eq:1.2a}). This illustrates that the absence of a singular contribution to $\chi_{\text{L}}$,
and the related fact that the maximally spin-polarized state is an exact eigenstate of the Heisenberg ferromagnet, is not
due to the absence of quantum fluctuations in the ground state, as is sometimes stated in the literature. Rather, it is due
to the fact that $\chi_{\text{L}}$ can be formulated as a correlation function of the magnon number. Quantum fluctuations
do exist in the ground state of a ferromagnet, and they affect correlation functions, such as $\Psi$, that can not be 
formulated entirely in terms of fluctuations of the magnon number. The same holds for the longitudinal susceptibility in
an antiferromagnet, see Sec.~\ref{subsec:II.B} below. We will come back to this point in Sec.~\ref{subsubsec:IV.B.2}.

\subsubsection{Singularities at $T>0$}
\label{subsubsec:II.A.2}

To find the behavior at nonzero temperature we perform the Matsubara frequency sum in Eq.~(\ref{eq:2.11}). This yields
\be
\chi_{\text{L}}^{(1)}(k,H) = \frac{-1}{V}\!\sum_{{\bm p},\sigma}
      \frac{n(\omega_{\bm p} + \mu H) - n(\omega_{{\bm p}-{\bm k}} + \mu H)}{\omega_{\bm p} - \omega_{{\bm p}-{\bm k}} + \sigma i\Omega_n},
\label{eq:2.15}
\ee
where $n(x) = 1/(e^{x/T}-1)$ is the Bose distribution function (we use units such that $\hbar = k_{\text B} = 1$),
and $\omega_{\bm p} = D{\bm p}^2$ is the ferromagnetic magnon frequency. 
We are interested in infrared singularities that arise from the small-momentum behavior of the integrand in 
Eq.~(\ref{eq:2.15}). Accordingly, to obtain the leading singular behavior as ${\bm k} \to 0$ for fixed $T$, we can
expand the Bose function, $n(x) \approx T/x$.\cite{classical_limit_footnote} 
At zero external frequency, $k = ({\bm k},i0)$, and zero external field we have
\be
\chi_{\text{L}}^{(1)}({\bm k},H=0) \approx \left(\frac{M_0}{\rho_s}\right)^2 \frac{2T}{V} \sum_{\bm p} \frac{1}{{\bm p}^2 ({\bm p}-{\bm k})^2}\ ,
\label{eq:2.16}
\ee
where we have used $D = \rho_s/M_0$.
Note that this leading contribution is necessarily linear in $T$, and that the wavenumber integral is a convolution
of two classical Goldstone modes, see Eq.~(\ref{eq:1.1}). 
In $d=3$ we find explicitly
\be
\chi_{\text{L}}^{(1)}({\bm k},H=0) = \frac{T}{4D^{3/2}\sqrt{\omega_{\bm k}}}\left[1 + O(\sqrt{\omega_{\bm k}/T})\right]\ (d=3);
\label{eq:2.17}
\ee
in generic dimensions $2<d<4$ the singularity is proportional to $T/\vert{\bm k}\vert^{4-d}$ with a $d$-dependent prefactor. 
For $d\leq 2$ the singular integral has a zero prefactor since $M_0=0$.
This result is valid for $\mu H \ll \omega_{\bm k} \ll T$.
The range of validity of Eq.~(\ref{eq:2.16}) thus shrinks with decreasing temperature. In the asymptotic low-temperature
limit in a vanishingly small field, i.e., for $\mu H \ll T \ll \omega_{\bm k}$, we find
\be
\chi_{\text{L}}^{(1)}({\bm k},H=0) = \frac{c_{\text{L}}}{\pi^2}\,\frac{T^{3/2}}{D^{3/2}\omega_{\bm k}}\left[1 + O(T/\omega_{\bm k})\right]\ (d=3),
\label{eq:2.18}
\ee
where $c_{\text{L}} = \sqrt{\pi/2}\,\zeta(3/2) \approx 2.395$, with $\zeta$ the Riemann zeta function.
For $T < \omega_{\bm k}$ the $T/\sqrt{\omega_{\bm k}}$ singularity thus crosses over to $T^{3/2}/\omega_{\bm k}$,
and for $T\to 0$ the prefactor of the singularity vanishes in agreement with Sec.~\ref{subsubsec:II.A.1}.

For $\omega_{\bm k} \ll \mu H \ll T$ 
an analogous consideration yields
\be
\chi_{\text{L}}^{(1)}({\bm k}\to 0,H) = \frac{T}{4\pi D^{3/2} (\mu H)^{1/2}}\,\left[1 + O(\sqrt{H/T})\right]\ ,
\label{eq:2.19}
\ee
and for $T \ll \omega_{\bm k}, \mu H$ the leading behavior is
\be
\chi_{\text{L}}^{(1)}({\bm k},H) = \frac{1}{2\pi^{3/2}}\,\frac{T^{3/2}}{D^{3/2}\omega_{\bm k}}\,e^{-\mu H/T}\ .
\label{eq:2.20}
\ee 
Both of these results are for $d=3$. Finally, for $\omega_{\bm k} \ll T \ll \mu H$ the result is proportional to
$T^{1/2} e^{-\mu H/T}$ with no singular dependence on $\omega_{\bm k}$.

\subsubsection{The dynamical structure factor}
\label{subsubsec:II.A.3}

Also of interest is the longitudinal part of the dynamical structure factor 
$S_{\text{L}}({\bm k},\omega) = (2/(1-e^{-\omega/T})) \chi_{\text{L}}''({\bm k},\omega)$,
with $\chi_{\text L}''$ the spectrum of the susceptibility $\chi_{\text{L}}$. From Eq.~(\ref{eq:2.15}) we find for the one-loop 
contribution\cite{Vaks_Larkin_Pikin_1967}
\bea
S_{\text{L}}^{(1)}({\bm k},\omega) &=& \frac{1}{1-e^{-\omega/T}}\,\frac{T}{4\pi D^{3/2} \sqrt{\omega_{\bm k}}}\,
\nonumber\\
&& \hskip - 10pt \times \ln\left(\frac{1 - e^{-(\omega + \omega_{\bm k})^2/4T\omega_{\bm k} - \mu H/T}}
     {1 - e^{-(\omega - \omega_{\bm k})^2/4T\omega_{\bm k} - \mu H/T}}\right)\ .\qquad\qquad
\label{eq:2.21}
\eea
The leading behavior for small ${\bm k}$, $\omega$, and $H$ for fixed $T$ is
\bea
S_{\text{L}}^{(1)}({\bm k},\omega) &\approx& \frac{T^2}{4\pi D^{3/2}\omega \sqrt{\omega_{\bm k}}}\,
\nonumber\\
&& \hskip - 20pt \times \ln\left(\frac{(\omega + \omega_{\bm k})^2/4T\omega_{\bm k} + \mu H/T}
     {(\omega - \omega_{\bm k})^2/4T\omega_{\bm k} + \mu H/T}\right)\ .\qquad
\label{eq:2.22}
\eea
As in the case of Eq.~(\ref{eq:2.16}), this is also what one obtains in the classical limit, $\hbar\to 0$ (see
also Ref.~\onlinecite{classical_limit_footnote}, and note that $\mu/\hbar$ is independent of $\hbar$).

The structure factor is shown in Fig.~\ref{fig:3} for several values of $T/\omega_{\bm k}$. 
Notable features are as follows: (1) There is a logarithmic singularity at $\omega = \pm\omega_{\bm k}$. This
leads to a broad feature, even for undamped magnons, whose width is independent of the
normalized temperature. (2) There is a marked decrease in the overall value of $S_{\text{L}}$ with
decreasing temperature, and (3) $S_{\text{L}}$ becomes strongly asymmetric at low temperature due to the
detailed-balance factor. A nonzero magnetic field removes the logarithmic singularity, and
even a rather small magnetic field substantially broadens the resonance feature, see
Fig.~\ref{fig:4}. We will further discuss the dynamical structure factor in Sec.~\ref{sec:IV}.

We also note that the minus first frequency moment of $\chi_{\text{L}}''$ yields the static susceptibility:
$\chi_{\text{L}}({\bm k}) = \int_{-\infty}^{\infty} d\omega\,\chi_{\text{L}}''({\bm k},\omega)/\pi\omega$.
Performing the frequency integral we recover the results given in Eqs.~(\ref{eq:2.17}) - (\ref{eq:2.20}).

\subsection{Quantum antiferromagnets}
\label{subsec:II.B}

We now consider quantum antiferromagnets, whose spin dynamics are very different from their ferromagnetic counterparts.
The \NLSM\ for an antiferromagnet can be written \cite{Fradkin_1991, Sachdev_1999}
\bse
\label{eqs:2.23}
\be
Z = \int {\cal D}[\hat{\bm n}]\,\delta(\hat{\bm n}^2(x) - 1)\ e^{-\int dx\, {\cal L}_{\text{AFM}}[\hat{\bm n}]}
\label{eq:2.23a}
\ee
with an action density
\bea
{\cal L}_{\text{AFM}}[\hat{\bm n}] &=&  \frac{\rho_s}{2}\Bigl[-\hat{\bm n}(x)\cdot {\bm\nabla}^2 \hat{\bm n}(x) 
\nonumber\\
&& \hskip 0pt + \frac{1}{c^2} \bigl(\partial_{\tau}\hat{\bm n}(x)  -i \mu {\bm H} \times \hat{\bm n}(x) \bigr)^2 \Bigr] \ .\qquad\qquad
\label{eq:2.23b}
\eea
\ese
Here $\hat{\bm n}(x)$ is the normalized staggered magnetization. It obeys
\bse
\label{eqs:2.24}
\be
\hat{\bm n}^2(x) \equiv 1
\label{eq:2.24a}
\ee
and we parameterize it as
\be
\hat{\bm n}(x) = \left(\pi_1(x),\pi_2(x),\sigma(x)\right)
\label{eq:2.24b}
\ee
\ese
in analogy to the ferromagnetic case. The physical staggered magnetization is ${\bm N}(x) = N_0\,\hat{\bm n}(x)$ with an amplitude
$N_0$. $\rho_s$ is the spin stiffness, $c$ is the spin-wave velocity and ${\bm H}$ is a homogeneous external magnetic field. Notice that in 
the absence of an external field the dynamics are given by a $(\partial_{\tau}\hat{\bm n})^2$ term, in contrast to the linear dependence 
on $\partial_{\tau}$ in the ferromagnetic case, Eq.~(\ref{eq:2.2b}).\cite{1d_footnote} 
Putting the external field
equal to zero, and proceeding as in the ferromagnetic case, we obtain a transverse Gaussian fluctuation action that is diagonal in
the $\pi_1$-$\pi_2$ basis:
\bse
\label{eqs:2.25}
\be
{\cal A}^{(2)}[\pi_1,\pi_2] = \frac{\rho_s}{2 c^2} \sum_k \sum_{i=1}^{2} \pi_i(k)\,\mu(k)\,\pi_i(-k)\ ,
\label{eq:2.25a}
\ee
with an eigenvalue
\be
\mu(k) = \omega_{\bm k}^2 - (i\Omega_n)^2\ .
\label{eq:2.25b}
\ee
\ese 
Here $\omega_{\bm k} = c\vert{\bm k}\vert$ is the antiferromagnetic magnon frequency. The one-loop contribution to the longitudinal susceptibility 
$\chi_{\text{L}}(x-y) = N_0^2 \langle \delta\sigma(x)\,\delta\sigma(y)\rangle$ 
now has the form
\be
\chi_{\text{L}}^{(1)}(k) = \left(\frac{N_0\,c^2}{\rho_s}\right)^2\,\frac{T}{V}\sum_p \frac{1}{\mu(p) \mu(p - k)}\ .
\label{eq:2.26}
\ee
Notice that this is the longitudinal order-parameter susceptibility, which describes the response to a staggered magnetic
field, rather than to a homogeneous one.

\subsubsection{The Goldstone-mode-induced singularity at $T=0$}
\label{subsubsec:II.B.1}

The one-loop contribution to the longitudinal susceptibility given by Eq.~(\ref{eq:2.26}) is still represented by the diagram 
shown in Fig.~\ref{fig:2}, but now the frequency integration at $T=0$ involves poles on either side of the real axis. The
frequency integral thus does not vanish, and we obtain
\bse
\label{eqs:2.27}
\bea
\chi_{\text{L}}^{(1)}({\bm k},i\Omega_n=0) &=&  \left(\frac{N_0\,c^2}{\rho_s}\right)^2\,\frac{1}{2V}\sum_p \frac{1}{\omega_{{\bm p}+{\bm k}/2}\,
   \omega_{{\bm p}-{\bm k}/2}}\,
\nonumber\\
&& \hskip 20pt \times \frac{1}{\omega_{{\bm p}+{\bm k}/2} + \omega_{{\bm p}-{\bm k}/2}}\ ,
\label{eq:2.27a}\\
\chi_{\text{L}}^{(1)}({\bm k}=0,i\Omega_n) &=&  \left(\frac{N_0\,c^2}{\rho_s}\right)^2\,\frac{1}{V}\sum_p 
\frac{1}{\omega_{\bm p}}\,\frac{1}{4\omega_{\bm p}^2 + \Omega_n^2}\ .
\nonumber\\
\label{eq:2.27b}
\eea
\ese
This yields the result expected from naive power counting, Eq.~(\ref{eq:1.2a}):
\bea
\chi_{\text{L}}^{(1)}({\bm k},i\Omega_n=0) &\propto& \vert{\bm k}\vert^{d-3}\ ,
\nonumber\\
\chi_{\text{L}}^{(1)}({\bm k}=0,i\Omega_n) &\propto& \vert\Omega_n\vert^{d-3}
\label{eq:2.28}
\eea
for $1<d<3$. 
In time space the latter result corresponds to a $1/t^{d-2}$ long-time tail, see Appendix \ref{app:B.1}.
The above derivation makes it clear that the striking difference between the 
behavior of this correlation function for ferromagnets and antiferromagnets, respectively, is a direct consequence
of the different spin dynamics in the two cases.

In $d=3$ the divergence is logarithmic. Calculating the prefactor we obtain, keeping only the leading terms,
\bse
\label{eqs:2.29}
\bea
\chi_{\text{L}}^{(1)}({\bm k},i0) &=& \frac{N_0^2 c}{8\pi^2\rho_s^2}\,\log(\omega_0/\omega_{\bm k})\ ,
\label{eq:2.29a}\\
\chi_{\text{L}}^{(1)}({\bm k}=0,i\Omega_n) &=&  \frac{N_0^2 c}{8\pi^2\rho_s^2}\,\log(\omega_0/\vert\Omega_n\vert)\ ,
\label{eq:2.29b}\\
\chi_{\text{L}}^{(1)}({\bm k}=0,i\Omega_n &\to& \Omega + i0) =  \frac{N_0^2 c}{8\pi^2\rho_s^2}\Bigl[\log(\omega_0/\vert\Omega\vert)
\nonumber\\
    &&\hskip 50pt + i\,\frac{\pi}{2}\,\sgn\Omega\Bigr]\ ,
\label{eq:2.29c}
\eea
\ese
where $\omega_0$ is an ultraviolet cutoff wave frequency. In $d=2$ the explicit result is 
\bse
\label{eqs:2.30}
\bea
\chi_{\text{L}}^{(1)}({\bm k},i0) &=& \frac{N_0^2 c^2}{8\rho_s^2}\,\frac{1}{\omega_{\bm k}}\ .
\label{eq:2.30a}\\
\chi_{\text{L}}^{(1)}({\bm k}=0,i\Omega_n) &=& \frac{N_0^2 c^2}{8\rho_s^2}\,\frac{1}{\vert\Omega_n\vert}\ ,
\label{eq:2.30b}\\
\chi_{\text{L}}^{(1)}({\bm k}=0,i\Omega_n \to \omega + i0) &=& \frac{N_0^2 c^2}{8\rho_s^2}\,\left[\frac{i}{\omega}+\pi\delta(\omega)\right]\ .
\nonumber\\
\label{eq:2.30c}
\eea
\ese
Note that in time space Eq.~(\ref{eq:2.30b}) implies a correlation function that does not decay for long times, but rather is constant,
see Appendix \ref{app:B.3}. We will get back to this in Sec.~\ref{sec:IV}.

\begin{figure*}[t]
\includegraphics[width=14cm]{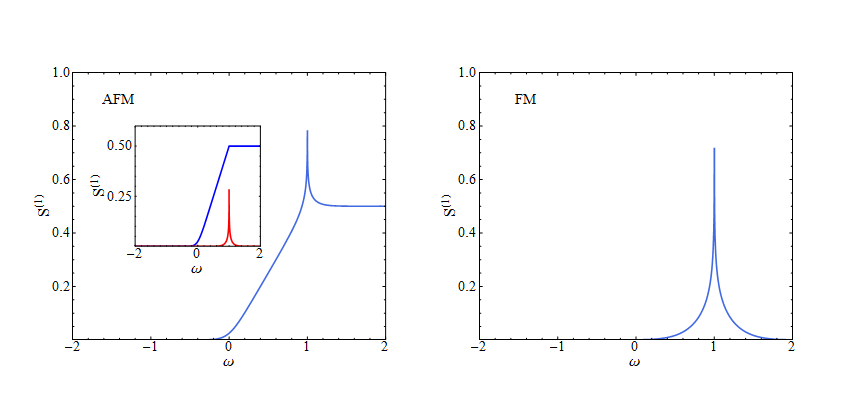}
\caption{The one-loop contribution to the longitudinal part of the dynamical structure factor for an antiferromagnet (left panel) 
              and a ferromagnet (right panel), 
              normalized by $N_0^2 c/4\pi\rho_s^2$ and as in Figs.~\ref{fig:4} and \ref{fig:3}, respectively, for $T/\omega_{\bm k} = 0.05$ 
              as functions of the frequency $\omega$ measured in units of $\omega_{\bm k}$. The inset in the left panel separately 
              shows the $T=0$ contribution to the antiferromagnetic structure factor (blue curve) and the contribution that vanishes
              as $T\to 0$ (red curve). The structure factor shown in the main panel is the sum of these two, see Eq.~(\ref{eq:2.36}).}
\label{fig:5}
\end{figure*}

\subsubsection{Singularities at $T>0$}
\label{subsubsec:II.B.2}

We now demonstrate that at a nonzero temperature we obtain the same result as in the ferromagnetic case. 
Performing the frequency summation in Eq.~(\ref{eq:2.26}) we obtain
\be
\chi_{\text{L}}^{(1)}(k) = \left(\frac{N_0\,c^2}{\rho_s}\right)^2 \frac{-1}{2V}\,\sum_{{\bm p},\sigma} \frac{1}{\omega_{\bm p}}\,\frac{n(\omega_{\bm p}) - n(-\omega_{\bm p})}
 {(\omega_{\bm p} + \sigma i\Omega_n)^2 - \omega_{{\bm p}+{\bm k}}^2}\ .
\label{eq:2.31}
\ee
The leading infrared behavior again comes from the small-momentum behavior of the integrand, so we approximate
$n(x) \approx T/x$. The resulting expression at zero external frequency can be rewritten to yield
\be
\chi_{\text{L}}^{(1)}({\bm k},i0) \approx  \left(\frac{N_0}{\rho_s}\right)^2\,\frac{T}{V} \sum_{\bm p} \frac{1}{{\bm p}^2 ({\bm p}-{\bm k})^2}\ .
\label{eq:2.32}
\ee
As in the ferromagnetic case, Eq.~(\ref{eq:2.16}), this indeed reproduces Eq.~(\ref{eq:1.1}). In $d=3$ we have explicitly
\bea
\chi_{\text{L}}^{(1)}({\bm k},i0) &=& \frac{N_0^2 c}{8\rho_s^2}\,\frac{T}{\omega_{\bm k}}\,\left[1 + O\bigl((\omega_{\bm k}/T)
               \log(\omega_0/\omega_{\bm k})\bigr)\right]
\nonumber\\
&& \hskip 70pt         \qquad (d=3)\ ,
\label{eq:2.33}
\eea
which is valid for $\omega_{\bm k} \ll T \ll \omega_0$. 
% For asymptotically low temperatures, it crosses over to Eq.~(\ref{eq:2.29a}), which is valid for $T \ll \omega_{\bm k}$.

Upon taking the $T\to 0$ limit in Eq.~(\ref{eq:2.31}), when $n(\omega_{\bm p}) - n(-\omega_{\bm p}) \to 1$,
we correctly recover the integrals given in Eqs.~(\ref{eqs:2.27}). In particular, Eq.~(\ref{eq:2.33}) crosses over
to Eq.~(\ref{eq:2.29a}), which is valid for $T \ll \omega_{\bm k}$.

\subsubsection{The dynamical structure factor}
\label{subsubsec:II.B.3}

Calculating the spectrum of the susceptibility from Eq.~(\ref{eq:2.31}) we obtain the one-loop contribution to the longitudinal part
of the dynamical structure factor. In $d=3$ we find
\be
S_{\text{L}}^{(1)}({\bm k},\omega) = \frac{N_0^2 c}{4\pi\rho_s^2}\,\frac{T/\omega_{\bm k}}{1 - e^{-\omega/T}}\,
     \ln \biggl( \frac{\sinh (\vert\omega_{\bm k} + \omega\vert/4T)}{\sinh (\vert\omega_{\bm k} - \omega\vert/4T)}\biggr)\ .
\label{eq:2.34}
\ee
It is illustrative to rewrite this as
\bea
S_{\text{L}}^{(1)}({\bm k},\omega) &=& \frac{N_0^2 c}{16\pi\rho_s^2}\,\frac{1}{1 - e^{-\omega/T}}\,\Biggl[
     \Bigl\vert 1 + \frac{\omega}{\omega_{\bm k}} \Bigr\vert - \Bigl\vert 1 - \frac{\omega}{\omega_{\bm k}}\Bigr\vert
\nonumber\\
&& + \frac{4T}{\omega_{\bm k}}\,\ln \left(\frac{1 - e^{-\vert\omega_{\bm k} + \omega\vert/2T}}{1 - e^{-\vert\omega_{\bm k} - \omega\vert/2T}}
          \right)\Biggr]\ .
\label{eq:2.35}
\eea
This separates $S_{\text{L}}^{(1)}$ into a contribution that survives the $T\to 0$ limit, and another one that is qualitatively
very similar to the corresponding result in the ferromagnetic case, see Eq.~(\ref{eq:2.21}). The former represents the quantum
fluctuations that are responsible for the singular behavior of $\delta \chi_{\text{L}}^{(1)}({\bm k})$ at $T=0$, and the latter
has again has a logarithmic singularity at the magnon resonance frequency $\omega = \omega_{\bm k}$. Note that the zero-temperature
contribution does not fall off as $\omega\to\infty$, but is constant. This statement is equivalent to the logarithmic divergence in
the static susceptibility: Calculating the minus first frequency moment of the spectrum 
$\chi_{\text{L}}''({\bm k},\omega) = (1 - e^{-\omega/T}) S_{\text{L}}({\bm k},\omega)/2$ in the limit $T\to 0$ we recover
Eq.~(\ref{eq:2.29a}). The difference between the antiferromagnetic and ferromagnetic cases becomes pronounced for 
temperatures $T \ll \omega_{\bm k}$; Fig.~\ref{fig:5} shows the respective results for $T/\omega_{\bm k} = 0.05$.

In the classical limit we have
\be
S_{\text{L}}^{(1)}({\bm k},\omega) = \frac{N_0^2 c}{4\pi\rho_s^2}\,\frac{T^2}{\omega\,\omega_{\bm k}}\,
      \ln \biggl( \frac{\omega_{\bm k} + \omega}{\omega_{\bm k} - \omega}\biggr)^2\ ,
\label{eq:2.36}
\ee
which is analogous to Eq.~(\ref{eq:2.22}).

In $d=2$ at $T=0$ the result is
\be
S_{\text{L}}^{(1)}({\bm k},\omega) = \frac{N_0^2 c^2}{4\rho_s^2}\,\Theta(\omega^2-\omega_{\bm k}^2)\,\frac{\Theta(\omega)}
   {\sqrt{\omega^2 - \omega_{\bm k}^2}}\ ,
\label{eq:2.37}
\ee
and calculating the minus first frequency moment recovers Eq.~(\ref{eq:2.30a}). For $T>0$ there is no long-range order in $d=2$.
% one finds a linear divergence of the longitudinal structure factor at the magnon resonance, $S_{\text{L}}^{(1)}\propto T/\vert \omega - \omega_{\bm k}\vert$.  However, this singularity has a zero prefactor since there is no long-range order in $d=2$ at $T>0$, and hence $N_0=0$. Similarly, the divergent momentum integral in Eq.~(\ref{eq:2.32}) has a zero prefactor; see also Ref.~\onlinecite{2d_footnote}.

\subsubsection{Quantum antiferromagnets in an external magnetic field}
\label{subsubsec:II.B.5}

So far we have considered the case of a vanishing external magnetic field. We now briefly discuss the effects of keeping the
field ${\bm H}$ in the action density, Eq.~(\ref{eq:2.23b}). The $({\bm H}\times{\hat{\bm n}})^2$ term in the action implies 
that in the ground state the order-parameter vector ${\bm n}$ is perpendicular to ${\bm H}$. Let ${\bm H}$ point in the
$x$-direction, ${\bm H}= (H,0,0)$, and we parameterize ${\bm n}$ as before in Eq.~(\ref{eq:2.24b}). The we find a Gaussian
action
\bse
\label{eqs:2.38}
\be
{\cal A}^{(2)}[\pi_1,\pi_2] = \frac{\rho_s}{2 c^2} \sum_k \sum_{i=1}^{2} \pi_i(k)\,\mu_i(k)\,\pi_i(-k)\ ,
\label{eq:2.38a}
\ee
where
\be
\mu_1(k) = \mu(k) + (\mu H)^2\quad,\quad \mu_2(k) = \mu(k)\ ,
\label{eq:2.38b}
\ee
\ese
with $\mu(k)$ from Eq.~(\ref{eq:2.27b}). Of the two Goldstone modes, one is thus unchanged, whereas the
other one acquires a mass. Equation (\ref{eq:2.26}) thus gets generalized to
\be
\chi_{\text{L}}^{(1)}(k) = \left(\frac{N_0\,c^2}{\rho_s}\right)^2\,\frac{T}{2V}\sum_p \sum_i \frac{1}{\mu_i(p) \mu_i(p - k)}\ ,
\label{eq:2.39}
\ee
and there is a singularity for ${\bm k}\to 0$ even for $H\neq 0$. At $T=0$ in $d=3$ we find, to leading logarithmic accuracy,
\bea
\chi_{\text{L}}^{(1)}({\bm k},i0) &=& \frac{N_0^2 c}{16\pi^2\rho_s^2} \biggl[\log\left(\frac{\omega_0}{\omega_{\bm k}}\right)
\nonumber\\
     && \hskip 0pt + \log\left(\frac{\omega_0}{\sqrt{\omega_{\bm k}^2 + (2\mu H)^2}}\right)\biggr].
\label{eq:2.40}
\eea
The corresponding result in $d=2$ is
\bse
\label{eqs:2.41}
\be
\chi_{\text{L}}^{(1)}({\bm k},i0) = \frac{N_0^2 c^2}{16\rho_s^2} \frac{1}{\omega_{\bm k}} 
     \left[1 + \frac{2}{\pi^2}\,g(\omega_{\bm k}/\sqrt{\omega_{\bm k}^2+(2\mu H)^2})\right].
\label{eq:2.41a}
\ee
where
\be
g(x) = \int_{-1}^{1} d\eta\,\frac{\ln(1+x\eta)}{\eta\sqrt{1-\eta^2}} = \begin{cases} \pi^2/2  & \text{for $x = 1$} \\
                                                                                                                             \pi x      & \text{for $x\to 0$}
                                                                                                      \end{cases}\ .
\label{eq:2.41b}
\ee
\ese
For $\mu H \ll \omega_{\bm k}$ we recover Eq.(\ref{eq:2.30a}); for $\mu H \gg \omega_{\bm k}$ we have
$\chi_{\text{L}}^{(1)}({\bm k},i0) \propto 1/H$.
Corresponding results are obtained for $\chi_{\text{L}}^{(1)}$ as a function of the frequency.

\section{Effects of damped ferromagnetic magnons}
\label{sec:III}

So far we have ignored the effects of damping on the magnons. In this section we will consider the effects of magnon damping on
the longitudinal susceptibility and the longitudinal dynamical structure factor in ferromagnets. We restrict ourselves to the ferromagnetic case,
where magnon damping has a qualitative effect. 

\subsection{Time-dependent Ginzburg-Landau theory}
\label{subsec:III.A}

We need to determine the effects of damping on the ferromagnetic Goldstone mode, Eq.~(\ref{eq:2.9a}). To this end
we use the standard time-dependent Ginzburg-Landau theory for a ferromagnet\cite{Landau_Lifshitz_1935, Ma_1976,
Hohenberg_Halperin_1977}
\bse
\label{eqs:3.1}
\bea
\partial_t{\bm M}({\bm x},t) &=& {\bm M}({\bm x},t)\times \frac{\delta S}{\delta{\bm M}({\bm x})}\biggr\vert_{{\bm M}({\bm x},t)}
\nonumber\\
                        && \hskip -20pt- \int d{\bm y}\,\Gamma({\bm x}-{\bm y})\,\frac{\delta S}{\delta{\bm M}({\bm y})}\biggr\vert_{{\bm M}({\bm y},t)}\ .
\label{eq:3.1a}
\eea
Here $\Gamma({\bm x})$ is the damping operator, which we will specify below, and $S$ is a
suitable action for the static magnetization ${\bm M}({\bm x})$. 
%Using a classical \NLSM\ we have, 
Very general considerations yield,
to linear order in ${\bm M}$,
\be
\delta S/\delta{\bm M}({\bm x}) = -(\rho_s/M_0^2) {\bm\nabla}^2{\bm M}({\bm x}) - \mu{\bm H}\ .
\label{eq:3.1b}
\ee
\ese
Here we use the same notation as in Sec.~\ref{sec:II} for the prefactor of the gradient-squared term.

We now use Eqs.~(\ref{eqs:3.1}) to calculate the linear response of the transverse magnetization components to the
external field ${\bm H}$, i.e., the transverse magnetic susceptibility $\chi_T$. The result is Eq.~(\ref{eq:2.4b}) with
the substitution $\Omega_n \to \Omega_n + \Gamma_{\bm k}\,{\bm k}^2\,\sgn(\Omega_n)$, where $\Gamma_{\bm k}$ is
the Fourier transform of $\Gamma({\bm x})$.
The one-loop contribution to the longitudinal susceptibility is still given by Eq.~(\ref{eq:2.11}), but with $\lambda_{\pm}$
replaced by
\be
\lambda_{\pm}({\bm k},i\Omega_n) = D{\bm k}^2 + \mu H \mp i\Omega_n \mp i\Gamma_{\bm k}\,{\bm k}^2\,\sgn(\Omega_n)\ .
\label{eq:3.2}
\ee
The $\sgn(\Omega_n)$ in the damping term follows from causality requirements.
In the absence of damping, $\Gamma_{\bm k} \equiv 0$, we recover the expressions given in Sec.~\ref{subsec:II.A}.

We expand the damping coefficient in the long-wavelength limit as
\be
\Gamma_{{\bm k}\to 0} = \gamma_0 + \gamma_2\, {\bm k}^2
\label{eq:3.3}
\ee
and distinguish between two physically distinct cases:\cite{Hohenberg_Halperin_1977} 
(1) A non-conserved order parameter, in which case $\gamma_0 > 0$, and
(2) a conserved order parameter, in which case $\gamma_0 = 0$. The former case is
realized, e.g., by magnetic impurities;\cite{Kohno_Tatara_Shibata_2006, Umetsu_Miura_Sakuma_2012}
the latter, by, e.g., damping by electron-magnon and/or magnon-magnon interactions at $T>0\,$\cite{Isoda_1990} 
or by nonmagnetic quenched disorder at any temperature, including 
$T=0.$\cite{Korenman_Prange_1972, Tserkovnyak_Hankiewicz_Vignale_2009, Umetsu_Miura_Sakuma_2012} 

\subsection{Non-Conserved order parameter}
\label{subsec:III.B}

We now perform the integral in Eq.~(\ref{eq:2.11}) with $\lambda_{\pm}$ given by Eq.~(\ref{eq:3.2}). For a non-conserved order
parameter, $\Gamma_{\bm p} = \gamma_0$, and again keeping only the leading terms, we find for $d=3$
\bse
\label{eqs:3.4}
\bea
\chi_{\text{L}}^{(1)}({\bm k}\to 0,i0) &=& \text{const.} - \frac{1}{32\pi}\,\frac{\gamma_0/\sqrt{D}}{\gamma_0^2 + D^2}\,\sqrt{\omega_{\bm k}}\ ,\qquad
\label{eq:3.4a}\\
\chi_{\text{L}}^{(1)}({\bm k} = 0,i\Omega_n) &=& - \frac{\gamma_0}{\pi^3D^{5/2}}\,f(\gamma_0/D)\,\vert\Omega_n\vert^{1/2}\ ,
\label{eq:3.4b}
\eea
\bea
\chi_{\text{L}}^{(1)}({\bm k} = 0,i\Omega_n \to \omega + i0) &=& \text{const.} \hskip 50pt
\nonumber\\
&& \hskip -120pt - \frac{\gamma_0}{\sqrt{2}\pi^3D^{5/2}}\,f(\gamma_0/D)\,[1 - i\, \sgn(\omega)]\,\vert\omega\vert^{1/2} \ .
\label{eq:3.4c}
\eea
The function $f$ can be expressed in terms of elementary functions; 
however, both the derivation and the result are
lengthy, see Appendix \ref{app:A}. Here we give only the power-series expansion for small damping, which reads
\be
f(x \to 0) = \frac{\sqrt{2}\pi}{5} + \frac{3\pi}{7\sqrt{2}}\,x + O(x^2)\ .
\label{eq:3.4c}
\ee
\ese
In $d=2$ there is a logarithmic singularity,
\bse
\label{eqs:3.5}
\bea
\chi_{\text{L}}^{(1)}({\bm k}\to 0,i0) &=& \frac{1}{2\pi^2}\,\frac{\gamma_0}{\gamma_0^2 + D^2}\,\ln(\omega_0/\omega_{\bm k})\ , \qquad\quad
\label{eq:3.5a}\\
\chi_{\text{L}}^{(1)}({\bm k} = 0,i\Omega_n) &=& \frac{1}{2\pi^2}\,\frac{\gamma_0}{D^2 + \gamma_0^2}\,\ln(\omega_0/\vert\Omega_n\vert)\ ,
\label{eq:3.5b}
\eea
\bea
\chi_{\text{L}}^{(1)}({\bm k} = 0,i\Omega_n \to \omega + i0) &=& \frac{1}{2\pi^2}\,\frac{\gamma_0}{D^2 + \gamma_0^2}\,
     \Bigl[\ln(\omega_0/\vert\omega\vert)
\nonumber\\
&& \hskip 20pt + i\,\frac{\pi}{2}\,\sgn\omega\Bigr]\ .
\label{eq:3.5c}
\eea
\ese
In generic dimensions the nonanalytic contribution is proportional to $\vert{\bm k}\vert^{d-2}$ and $\Omega_n^{(d-2)/2}$,
respectively. In time space the latter corresponds to a $1/t^{d/2}$ long-time tail, see Appendix \ref{app:B.1}.

\subsection{Conserved order parameter}
\label{subsec:III.C}

For a conserved order parameter, $\Gamma_{\bm p} = \gamma_2 {\bm p}^2$, the calculations are analogous but
more involved and we give the results only to linear order in $\gamma_2$. For $d=3$ we find
\bse
\label{eqs:3.6}
\bea
\chi_{\text{L}}^{(1)}({\bm k}\to 0,i0) &=& \text{const.} + O({\bm k}^2) 
\nonumber\\
&& \hskip 10pt + \frac{\gamma_2\left[1+O(\gamma_2^2)\right]}{64\pi D^{7/2}}\,\omega_{\bm k}^{3/2},
\label{eq:3.6a}
\eea
\be
\chi_{\text{L}}^{(1)}({\bm k} = 0,i\Omega_n) =  \text{const.} - \frac{\sqrt{2}\, \gamma_2\left[1+O(\gamma_2^2)\right]}{7\pi^2 D^{7/2}}\,
   \vert\Omega_n\vert^{3/2}\ ,
\label{eq:3.6b}
\ee
\bea
\chi_{\text{L}}^{(1)}({\bm k} = 0,i\Omega_n &\to& \omega + i0) = \text{const.} \hskip 50pt
\nonumber\\
&& \hskip -60pt + \frac{\gamma_2\left[1+O(\gamma_2^2)\right]}{7 \pi^2 D^{7/2}}\,[1 + i\, \sgn(\omega)]\,\vert\omega\vert^{3/2} \ .
\label{eq:3.6c}
\eea
\ese
In $d=2$ the leading singularity is
\bse
\label{eqs:3.7}
\be
\chi_{\text{L}}^{(1)}({\bm k}\to 0,i0) = \text{const.} - \frac{\gamma_2\left[1+O(\gamma_2^2)\right]}{48\pi^2 D^3}\,
\omega_{\bm k}\ln(\omega_0/\omega_{\bm k})\ ,
\label{eq:3.7a}
\ee
\be
\chi_{\text{L}}^{(1)}({\bm k} = 0,i\Omega_n) = \text{const.} -  \frac{\gamma_2\left[1+O(\gamma_2^2)\right]}{6\pi D^3}\,\vert\Omega_n\vert\ .
\label{eq:3.7b}
\ee
\be
\chi_{\text{L}}^{(1)}({\bm k} = 0,i\Omega_n\to\omega + i0) = \text{const.} +  \frac{i\gamma_2\left[1+O(\gamma_2^2)\right]}{12\pi D^3}\,\omega\ .
\label{eq:3.7c}
\ee
\ese 
In generic dimensions the nonanalytic contribution is proportional to $\vert{\bm k}\vert^{d}$ and $\Omega_n^{d/2}$,
respectively. In time space this corresponds to a $1/t^{(d+2)/2}$ long-time tail, see Appendix \ref{app:B.1}.

\section{Discussion}
\label{sec:IV}

In this final section we give a summary of our results and conclude with a discussion of various physical points
that underly them.

\subsection{Summary}
\label{subsec:IV.A}

In summary, we have investigated the coupling of magnons in quantum ferromagnets and antiferromagnets to 
other correlation functions, in particular the longitudinal susceptibility and the longitudinal part of the dynamical structure factor. In
the case of ferromagnets with undamped magnons the longitudinal susceptibility vanishes at $T=0$. In $d=3$, 
and in the absence of an external magnetic field, 
an interpolating expression that correctly describes the leading behavior for both
$T>\omega_{\bm k}$ and $T<\omega_{\bm k}$ is
\be
\chi_{\text{L}}^{(1)}({\bm k},H=0) = \frac{T}{4D^{3/2}\sqrt{\omega_{\bm k}}}\,\frac{1}{1+(\pi^2/c_{\text{L}})\sqrt{\omega_{\bm k}/T}}\ ,
\label{eq:4.1}
\ee
where $\omega_{\bm k} = D{\bm k}^2$ is the ferromagnetic magnon frequency and $c_{\text{L}}$ is the constant given 
after Eq.~(\ref{eq:2.18}). For $T>\omega_{\bm k}$ one has the
classical $1/\vert{\bm k}\vert$ singularity, Eq.~(\ref{eq:2.17}), whereas for $T<\omega_{\bm k}$ $\chi_{\text{L}}$ vanishes as
$T^{3/2}$, Eq.~(\ref{eq:2.18}). For a quantum antiferromagnet, the corresponding interpolating expression is
\be
\chi_{\text{L}}^{(1)}({\bm k},i0) = \frac{N_0^2 c}{8\rho_s^2}\,\frac{T}{\omega_{\bm k}}\,\left[1 + (\omega_{\bm k}/\pi^2 T)
               \log(\omega_0/\omega_{\bm k})\right]\ ,
\label{eq:4.2}
\ee
see Eqs.~(\ref{eq:2.29a}) and (\ref{eq:2.33}). Here $\omega_{\bm k} = c\vert{\bm k}\vert$ is the antiferromagnetic magnon
frequency. This reflects the expected scaling behavior, viz., $1/\vert{\bm k}\vert$ for high temperature, and $\ln \vert{\bm k}\vert$
for low temperature. Similarly, the longitudinal dynamical structure factor for a ferromagnet vanishes at $T=0$, see Eq.~(\ref{eq:2.22})
and Fig.~\ref{fig:3}, whereas in the antiferromagnetic case there is a nonvanishing contribution even at $T=0$, see
Eq.~(\ref{eq:2.35}) and Fig.~\ref{fig:5}. Quenched disorder introduces additional fluctuations, leads to magnon damping, and 
qualitatively changes the ferromagnetic results. Magnetic impurities, which lead to a non-conserved magnetization,
results in the longitudinal susceptibility scaling as $\vert{\bm k}\vert^{d-2}$, where the zero exponent in $d=2$
signifies a logarithmic divergency, see Sec.~\ref{subsec:III.B}. Non-magnetic disorder leads to a weaker scaling
behavior, $\vert{\bm k}\vert^d$, see Sec.~\ref{subsec:III.C}. For $T>0$ the longitudinal dynamical structure factor has a logarithmic 
singularity at the magnon frequency in both ferromagnets and antiferromagnets.

\subsection{Discussion}
\label{subsec:IV.B}

We conclude with a discussion of various physical points raised by our results.

\subsubsection{Predictions for experiments}
\label{subsubsec:IV.B.1}

\paragraph{a) Longitudinal susceptibility and dynamical structure factor:}
\label{par:IV.B.1.a}

The classical singularity of $\chi_{\text{L}}$ in the ferromagnetic case as a function of an external magnetic field,
Eq.~(\ref{eq:2.19}), has been observed by K{\"o}tzler et al.\cite{Kotzler_et_al_1994} The theoretical prediction is
that in the limit of low temperatures, $T \ll \mu H$,  $\chi_{\text{L}}$ becomes exponentially
small, see Eq.~(\ref{eq:2.20}) and the paragraph following it.

A remarkable feature in the longitudinal dynamical structure factor is the logarithmic singularity at the magnon resonance
frequency, see Eqs.~(\ref{eq:2.21}) and (\ref{eq:2.34}), and Fig.~\ref{fig:3}. In a clean system at low 
temperature the magnon damping is very weak, and the magnon peaks in the transverse dynamical susceptibility
are very narrow. The longitudinal susceptibility or structure factor, by contrast, shows an intrinsically broad feature
at the magnon frequency. Even a rather small magnetic field substantially broadens and suppresses this feature,
see Fig.~\ref{fig:4}.

\bigskip\par
\paragraph{b) Other correlation functions:}
\label{par:IV.B.1.b}

We stress again that the behavior of the longitudinal susceptibility is not generic,
but rather restricted to a class of correlation functions that can be expressed entirely in terms of magnon number
fluctuations. Other correlation functions do show the expected $\omega^{(d-2)/2}$ frequency scaling, see the
example in Sec.~\ref{subsubsec:II.A.2}. 

An example of a correlation function that belongs to the same class as the longitudinal susceptibility is the electrical 
conductivity in a metallic quantum ferromagnet; they both share the same scaling behavior. This implies that undamped magnons
do {\em not} lead to an $\omega^{(d-2)/2}$ frequency dependence of the conductivity at $T=0$, or a $\ln\omega$
singularity in $d=2$. The latter conclusion was reached correctly in Ref.~\onlinecite{Kirkpatrick_Belitz_2000}, but
a sign error incorrectly led to the prediction of an $\omega^{(d-2)/2}$ nonanalyticity in $d>2$. A corrected analysis
of the conductivity in itinerant ferromagnets will be given elsewhere.\cite{us_tbp}

\subsubsection{Comments on the results for ferromagnets}
\label{subsubsec:IV.B.2}

\paragraph{a) Fluctuations and entanglement entropy:} Let us come back to the issue of fluctuations in the ground state
of a ferromagnet, see the remarks at the end of Sec.~\ref{subsubsec:II.A.2}. A global measurement of fluctuations in a
system is given by the entanglement entropy, defined as the von Neumann entropy of a subsystem of linear size $L$.
At zero temperature the entropy vanishes in the thermodyamic limit, and for $L\to\infty$ it grows more slowly than the
volume $L^d$. In systems that do not contain a Fermi surface the leading contribution is in general given by an ``area-law'' term
that grows as $L^{d-1}$;\cite{Eisert_Cramer_Plenio_2010} this term is due to short-range entanglement and has a 
non-universal prefactor. The leading universal contribution, which is a measure of long-range fluctuations, in systems
with Goldstone modes grows as $\ln L$. This is true for both quantum ferromagnets\cite{Popkov_Salerno_2005, 
Ding_Bonesteel_Yang_2008} and antiferromagets\cite{Song_et_al_2011, Metlitski_Grover_2015, Misguich_Pasquier_Oshikawa_2016} 
for $d=2,3$, although the area-law term is missing in the former.\cite{Ding_Bonesteel_Yang_2008}
This is another indication that fluctuations exist in the ferromagnetic ground state, although they may or may 
not be probed by a specific correlation function.

In metallic magnets, and more generally in systems with a Fermi surface, there is an area-law term with a multiplicative
logarithm that is due to long-range fluctuations in the fermionic degrees of freedom. This is one of many indications of
fundamental differences between metallic and insulating magnets. We briefly discuss some of these next.

\bigskip\par
\paragraph{b) Spin models vs. itinerant magnets:} There are important differences between the fluctuations in quantum
ferromagnets vs. antiferromagnets, the qualitatively similar universal parts of the entanglement entropies discussed above notwithstanding.
For instance, a spin model for a Heisenberg ferromagnet, with Hamiltonian $H = J\sum_{<ij>}{\bm{\sigma}}_i\cdot{\bm{\sigma}}_j$
with $J<0$, has no quantum phase transition as a function of $J$, since the ground state is fully spin-polarized for any $J<0$.
In this sense the quantum fluctuations in a ferromagnet, while present, are weaker than those in a quantum antiferromagnet.
This argument must survive nonmagnetic quenched disorder, which makes $J$ a random function of spatial position, as
long as the distribution of $J$ is restricted to negative values, since the spins will still be locally maximally polarized. Since
the physical reason for the absence of a nonanalyticity in $\chi_{\text{L}}$ is the same as that for the absence of a quantum
phase transition, it follows that nonmagnetic disorder with a such restricted distribution cannot lead to magnon damping;
the damping coefficient $\gamma_2$ in Sec.~\ref{subsec:III.C} must vanish at $T=0$. These considerations raise 
interesting questions about the strength of quantum fluctuations, as well as ways to measure them, see, e.g.,
Refs.~\onlinecite{Song_et_al_2011, Malpetti_Roscilde_2016}.

These aspects change qualitatively in a metallic ferromagnet, and in particular in an itinerant one: The coupling of the
magnetic degrees of freedom to the fermionic ones leads to a large increase in fluctuations. As a result, the
entanglement entropy has an area-law term multiplied by a logarithm, as is typical for systems with a Fermi surface,
and as a function of the exchange coupling there is a quantum phase transition, as first described by Stoner.\cite{Stoner_1938}
It is also likely that the presence of nonmagnetic disorder leads to magnon damping irrespective of the shape of the
disorder distribution. For a recent review of metallic ferromagnets, see Ref.~\onlinecite{Brando_et_al_2016a}. An explicit
discussion of $\chi_{\text{L}}$ and related correlation functions in a model of itinerant ferromagnets will be given
elsewhere.\cite{us_tbp}

\bigskip\par
\paragraph{c) Effects of quenched disorder:} We now discuss the fact that quenched disorder, and the resulting damping
of the magnons, leads to a 
nonanalyticity in $\chi_{\text{L}}$, and demonstrate that the result is consistent with scaling and renormalization-group
considerations and is indeed asymptotically exact as far as the exponent of the nonanalyticity is concerned. 

First of all, we recall that the absence of a nonanalyticity for systems with undamped magnons is due to the absence
of fluctuations that couple to the longitudinal magnetization fluctuations. Disorder introduces additional fluctuations,
which makes it plausible that it will lead to a nonanalyticity. Furthermore, magnetic disorder, which couples directly to the
order parameter, will have a stronger effect than nonmagnetic disorder, and thus result in a stronger singularity. The 
results in Secs.~\ref{subsec:III.B} and \ref{subsec:III.C} thus are physically plausible.

In order to deduce the explicit results from general arguments, we consider the Gaussian action written in the form
of Eq.~(\ref{eqs:2.3}) or (\ref{eq:2.6}), and add damping according to the prescription given above Eq.~(\ref{eq:3.2}).
In a schematic notation that shows only what is necessary for power counting the Gaussian action then takes the form
\be
{\cal A}^{(2)} = \int dx\,\pi(x)\left[D\partial_{\bm x}^2 + \partial_{\tau} + H + \gamma_n\,\partial_{\bm x}^{n+2}\right]\pi(x)\ .
\label{eq:4.3}
\ee
Here $n=0$ and $n=2$ correspond to the cases of a non-conserved and conserved order parameter, respectively. 
Additional terms in the action fall into two classes: (1) Gaussian with additional gradients, with the leading terms
of the form
\bse
\label{eqs:4.4}
\be
\delta{\cal A}^{(2)} = \int dx\,\partial_{\bm x}^4\,\pi^2(x)\ ,
\label{eq:4.4a}
\ee
or equivalent in terms of scale dimensions. (2) Of higher order in $\pi$, with the leading terms of the form
\be
\delta{\cal A}^{(4)} = \int dx\,\partial_{\bm x}^2\,\pi^4(x)
\label{eq:4.4b}
\ee
\ese
or equivalent. 

We now sketch a renormalization-group analysis of this action. In doing so, we follow a scheme pioneered by
Ma,\cite{Ma_1976} see also Refs.~\onlinecite{Belitz_Kirkpatrick_2014} and \onlinecite{Belitz_Kirkpatrick_1997} for applications
of this scheme in different contexts. We assign scale dimensions $[L] = -1$ and $[\tau] = -2$ to lengths and imaginary times,
respectively. Then there is a stable Gaussian fixed point where $\pi$ has a scale dimension $[\pi(x)] = d/2$.
In Fourier space this corresponds to $[\pi(k)] = -1$. We thus have $\langle\pi(k)\pi(-k)\rangle \sim 1/{\bm k}^2 \sim 1/\Omega_n$.
This scaling behavior describes the magnons, see Eqs.~(\ref{eqs:2.4}), and the Gaussian fixed point describes
the ordered phase where the symmetry is broken. The field $H$ is relevant with respect to this fixed point with
a scale dimension $[H] = 2$. For a non-conserved order parameter the damping coefficient $\gamma_0$ is
dimensionless, $[\gamma_0] = 0$, and the damping term is part of the fixed-point Hamiltonian. The free-energy
density $f$, the magnetization $m$, and the scaling part $\delta\chi_{\text{L}}$ of the longitudinal susceptibility 
$\chi_{\text{L}} = \partial m/\partial H$ then have scale dimensions $[f] = d-2$, $[m] = d$, and $[\delta\chi_{\text{L}}] = d-2$
respectively. For the latter this implies a homogeneity law
\be
\delta\chi_{\text{L}}({\bm k},i\Omega_n) = b^{2-d}\,F_{\chi}({\bm k} b,i\Omega_n b^2,\gamma_0)\ ,
\label{eq:4.5}
\ee
with $b$ an arbitrary length rescaling factor and $F_{\chi}$ a scaling function. The latter has the property 
$F_{\chi}(x,y,\gamma_0\to 0) = 0(\gamma_0)$, as we have discussed in the main part of this paper. We thus obtain the scaling behavior
\be
\delta\chi_{\text{L}}({\bm k},i\Omega_n) \sim \gamma_0\vert{\bm k}\vert^{d-2} \sim \gamma_0\vert\Omega_n\vert^{(d-2)/2}\ ,
\label{eq:4.6}
\ee
in agreement with Sec.~\ref{subsec:III.B}. The leading correction terms to the fixed-point action are irrelevant by
power counting, with scale dimensions $-2$ for the operator in Eq.~(\ref{eq:4.4a}) and $-d$ for the one in Eq.~(\ref{eq:4.4b}),
respectively. These arguments show that the one-loop results obtained in Sec.~\ref{subsec:III.B} are exact as
far as the exponents are concerned; higher terms in the loop expansion will change the prefactor of the nonanalyticity,
but not the power.

In the case of a conserved order parameter the damping term is not part of the fixed-point action; it is an irrelevant
operator with a scale dimension $[\gamma_2] = -2$ which is the same as the least irrelevant operators represented by,
e.g., Eq.~(\ref{eq:4.4a}). The homogeneity equation for $\delta\chi_{\text{L}}$ now reads
\bse
\label{eqs:4.7b}
\be
\delta\chi_{\text{L}}({\bm k},i\Omega_n) = b^{2-d}\,F_{\chi}({\bm k} b,i\Omega_n b^2,\gamma_2 b^{-2})\ ,
\label{eq:4.7a}
\ee
where we do not show the other irrelevant operators. Even though $\gamma_2$ is irrelevant, the scaling function still vanishes 
for $\gamma_2 = 0$, and we obtain, to linear order in $\gamma_2$,
\be
\delta\chi_{\text{L}}({\bm k},i\Omega_n) = b^{-d}\,\gamma_2\,{\tilde F}_{\chi}({\bm k} b,i\Omega_n b^2)\ ,
\label{eq:4.7a}
\ee
\ese
with ${\tilde F}$ another scaling function. This yields
\be
\delta\chi_{\text{L}}({\bm k},i\Omega_n) \sim \gamma_2\vert{\bm k}\vert^{d} \sim \gamma_2\vert\Omega_n\vert^{d/2}\ ,
\label{eq:4.8}
\ee
in agreement with Sec.~\ref{subsec:III.C}. Again, this is the exact leading scaling behavior.
\bigskip\par
\paragraph{d) Magnon damping:} An interesting aspect of ferromagnetic magnons is that these excitations cannot be
overdamped, irrespective of the magnitude of the damping coefficient. Consider Eq.~(\ref{eq:2.4b}) with 
$\Omega_n \to \Omega_n + \Gamma_{\bm k}\,{\bm k}^2\,\sgn(\Omega_n)$. The poles of $\chi_{\text{T}}({\bm k},z)$ with
$z$ the complex frequency, always have a real part given by $\pm\omega_{\bm k}$, independent of $\Gamma_{\bm k}$. 
This is in contrast to a damped harmonic oscillator, where the resonance frequency has no real part if the damping
coefficient is larger than a threshold value, and also to sound waves in fluids,\cite{Forster_1975} antiferromagnetic magnons, see 
Eqs.~(\ref{eqs:2.25}) with a damping coefficient added, and helimagnons in helical magnets,\cite{Belitz_Kirkpatrick_Rosch_2006a} 
all of which have the same structure as a damped harmonic oscillator.

% $\bullet$ TRK said: In the fluid case you basically couple a momentum to a density; and in the AFM case you couple an AFM order parameter to a homogeneous magnetization; note too, in the AFM case with a non conserved homogeneous magnetization things seem very, very different; i think the modes are then not propagating; one a kinetic mode; and one a diffusive mode.

\subsubsection{Comments on correlation functions that do not decay}
\label{subsubsec:IV.B.3}

We finally discuss the physical meaning of the constant long-time behavior implied 
by Eq.~(\ref{eq:2.30b}), see Eq.~(\ref{eq:B.17}). Let $T_{\text{max}}$ be the maximum time scale, which can be, e.g., the total duration 
of the experiment, or $L$ divided by the relevant characteristic velocity. $\chi_{\text{L}}$ then depends on two times, $t_1$ and $t_2$. 
As long as $t_1$, $t_2$, and $\vert t_1 - t_2\vert$ all are small compared to $T_{\text{max}}$, $\chi_{\text{L}}$ will not decay if 
$\vert t_1 - t_2\vert$ increases. In position space, by contrast, $\chi_{\text{L}}$ does decay, but only as a power: The $1/\vert{\bm k}\vert$ 
divergence in the $2$-$d$ quantum antiferromagnet, which is the same the one in a $3$-$d$ classical magnet, Eq.~(\ref{eq:1.1}), implies
that in real space the correlation function decays as $1/r$. For a general discussion of power-law decays of correlation functions, see, e.g.,
Ref.~\onlinecite{Belitz_Kirkpatrick_Vojta_2005}.

These results are examples of an effect that can be even stronger: In classical non-equilibrium fluids, and in Fermi liquids even in equilibrium, 
there are correlation functions that {\em increase} with increasing length or time scales in a well-defined sense, see 
Refs.~\onlinecite{Ortiz_Sengers_2007,  Kirkpatrick_Cohen_Dorfman_1982b, Kirkpatrick_Belitz_2016a}.

% $\bullet$ A different topic related to QAFMs: Hasselmann et al (2007) claim that there is something wrong with the AFM \NLSM\ in a magnetic field. They say it misses a quartic coupling, and fixing this leads to a magnetic-field dependence of the spin-wave stiffness coefficient $c$. We are not convinced that's really true; Sachdev's derivation of the \NLS \ in his book looks pretty convincing. And in any case this is not of crucial importance for anything we are doing. An interesting aside: The first instance of the \NLSM\ with a field I can find is D.S. Fisher (1989), but he does not give a derivation, and his references don't seem to talk about a field. So the derivation in Sachdev's book is the only one I can find.

\acknowledgments
We thank Max Metlitski for discussions. This work was supported by the NSF under Grants No. DMR-1401410 and No. DMR-1401449. 
Part of this work was performed at the Aspen Center for Physics, which is supported by the NSF under Grant No. PHY-1066293.

\appendix
\section{Frequency dependence of $\chi_{\text{L}}$ due to damped ferromagnetic magnons}
\label{app:A}

Here we sketch the derivation of Eq.~(\ref{eq:3.4b}) and give the full expression for the function $f$. Performing the
frequency sum in Eq.~(\ref{eq:2.15}) at $T=0$, with $\lambda_{\pm}$ given by Eq.~(\ref{eq:3.2}), we find
\begin{widetext}
\bea
\chi_{\text{L}}({\bm k}=0,i\Omega_n) &=& \frac{2}{\pi V}\sum_{\bm p} \Gamma_{\bm p}\,{\bm p}^2 \int_{\omega_{\bm p}^2}^{\infty} dx\ 
\frac{1}{x + (\Gamma_{\bm p}\,{\bm p}^2)^2}\,\frac{1}{x + (\Gamma_{\bm p}\,{\bm p}^2 + \Omega_n)^2}
\nonumber\\
&=& \frac{2}{\pi V}\sum_{\bm p} \Gamma_{\bm p}\,{\bm p}^2\,\frac{1}{\Omega_n^2 + 2\Omega_n \Gamma_{\bm p}\,{\bm p}^2}
        \ln\left(1 + \frac{\Omega_n^2 + 2\Omega_n \Gamma_{\bm p}\,{\bm p}^2}{\omega_{\bm p}^2 + (\Gamma_{\bm p}\,{\bm p}^2)^2}\right)     
\nonumber\\
&=& \frac{2}{\pi} \int_0^1 d\alpha\,\frac{1}{V}\sum_{\bm p} \frac{\Gamma_{\bm p}\,{\bm p}^2}{\alpha\Omega_n^2 + 2\alpha\Omega_n 
          \Gamma_{\bm p}\,{\bm p}^2 + \omega_{\bm p}^2 + (\Gamma_{\bm p}\,{\bm p}^2)^2}\ ,
\label{eq:A.1}
\eea
where in the last line we have expressed the logarithm in terms of an auxiliary integral. This procedure is also useful for
deriving the prefactors of the nonanalytic wave-number dependence at $\Omega_n=0$ that are given in Eqs.~(\ref{eq:3.4a}) and (\ref{eq:3.6a}).

We now consider the case of a non-conserved order parameter, $\Gamma_{\bm p} = \gamma_0$.
Splitting off the constant contribution
at $\Omega_n = 0$ in $d=3$, and scaling out the frequency, we obtain Eq.~(\ref{eq:3.4b}) with the function $f$ given by
\be
f(x) = \frac{1}{(1+x^2)^2} \int_0^1 d\alpha\,\alpha \int_0^{\infty} dy\,\frac{1+ 2 x y^2}{y^4 + 2 y^2 \alpha x/(1+x^2) + \alpha/(1+x^2)}\ .
\label{eq:A.2}
\ee
The integration over $y$ can now be easily performed, and the final integral over $\alpha$ can be expressed in terms of algebraic and 
inverse hyperbolic functions. We find
\be
f(x) = \frac{\pi}{6\sqrt{2}}\,\frac{1}{x^{5/2}(1+x^2)^{3/2}} \biggl\{\left[ 3 + 7 x^2 - 2x\sqrt{1+x^2}\right] \sqrt{x^2 + x\sqrt{1+x^2}}
- 3 (1+x^2)^{3/2} \sinh^{-1}(\sqrt{x}/(1+x^2)^{1/4})\biggr\}\ .
\label{eq:A.3}
\ee
\end{widetext}
An expansion for $x\to 0$ yields Eq.~(\ref{eq:3.4c}). In $d=2$ the logarithmic singularity is the leading term, and from
Eq.~(\ref{eq:A.1}) one readily obtains Eq.~(\ref{eq:3.5b}).

For a conserved order parameter, $\Gamma_{\bm p} = \gamma_2\,{\bm p}^2$, the integrals are more involved, but to linear order
in $\gamma_2$ one easily obtains Eqs.~(\ref{eq:3.6b}) and (\ref{eq:3.7b}) from Eq.~(\ref{eq:A.1}).

\section{Causal functions, and long-time tails}
\label{app:B}

Here we list, without proofs, some properties of the class of causal functions that the longitudinal susceptibility belongs to.
For general properties of causal functions see, e.g., Ref.~\onlinecite{Forster_1975}. For derivations of the long-time tails
see, e.g., Ref.~\onlinecite{Lighthill_1958}.

\subsection{Non-integer powers}
\label{app:B.1}

Consider a causal function $\chi$ of complex frequency $z$ that behaves, for $z\to 0$, as
\be
\chi(z) = \frac{1}{\cos(\alpha\pi/2)}\,\left[z^{\alpha} + (-z)^{\alpha}\right]\ ,
\label{eq:B.1}
\ee
with $\alpha$ real and not integer. Here and in what follows we consider even functions of $z$, since the
magnetic susceptibility has that property. We also give the asymptotic small-frequency, or long-time, behavior
only; for $z\to\infty$ $\chi$, or any causal function, must vanish. On the imaginary axis $\chi$ then takes the values
\be
\chi(i\Omega_n) = \vert\Omega_n\vert^{\alpha}\ ,
\label{eq:B.2}
\ee
and the spectrum $\chi''$ and the reactive part $\chi'$, respectively, of $\chi$ read
\bse
\label{eqs:B.3}
\bea
\chi''(\omega) &=& -\sin(\pi\alpha/2)\,\vert\omega\vert^{\alpha}\,\sgn\omega\ ,
\label{eq:B.3a}\\
\chi'(\omega) &=& \ \ \,  \cos(\pi\alpha/2)\,\vert\omega\vert^{\alpha}\ .
\label{eq:B.3b}
\eea
\ese
The real-time behavior of $\chi$ is given by the Fourier transform of $\chi''(\omega)$,
\be
\chi(t) = \int_{-\infty}^{\infty} \frac{d\omega}{\pi}\,e^{-i\omega t}\,\chi''(\omega)\ .
\label{eq:B.4}
\ee
In the long-time limit the Hardy-Littlewood 
tauberian theorem yields a long-time tail:
\be
\chi(t\to\infty) = i\,\frac{\Gamma(\alpha+1)}{\pi}\,\sin(\alpha\pi)\,\frac{1}{\vert t\vert^{\alpha+1}}
\label{eq:B.5}
\ee
The ferromagnet with damped magnons in $d=3$ is an example of this behavior, with $\alpha = 1/2$
and $\alpha = 3/2$ for a non-conserved and a conserved order parameter, respectively,
see Secs.~\ref{subsec:III.B} and \ref{subsec:III.C}. It is also realized by both ferromagnets and antiferromagnets
in generic dimensions.

\subsection{Even powers}
\label{app:B.2}

Now consider
\be
\chi(z) = \frac{(-)^m}{2}\,z^{2m}\left[\ln z + \ln(-z)\right]\ ,
\label{eq:B.6}
\ee
with $m$ integer. On the imaginary axis this yields
\be
\chi(i\Omega_n) = \vert\Omega_n\vert^{2m}\,\ln\vert\Omega_n\vert\ .
\label{eq:B.7}
\ee
The spectrum and the reactive part are
\bse
\label{eqs:B.8}
\bea
\chi''(\omega) &=& \frac{(-)^{m+1}\pi}{2}\,\omega^{2m}\,\sgn\omega\ ,
\label{eq:B.8a}\\
\chi'(\omega) &=& (-)^m\,\omega^{2m} \ln\vert\omega\vert\ ,
\label{eq:B.8b}
\eea
\ese
and the long-time behavior is
\be
\chi(t\to\infty) = i\, \frac{(2m)!}{\vert t\vert^{2m+1}}\ .
\label{eq:B.9}
\ee
Examples of this behavior are the antiferromagnet in $d=3$, Sec.~\ref{subsubsec:II.B.1},  and the ferromagnet in $d=2$ 
with a non-conserved order parameter, Sec.~\ref{subsec:III.B}.

\subsection{Odd powers}
\label{app:B.3}

Finally, consider
\be
\chi(z) = \frac{(-)^{m+1}}{\pi}\,z^{2m+1}\,\left[\ln z - \ln(-z)\right]\ ,
\label{eq:B.10}
\ee
with $m$ integer, which leads to
\be
\chi(i\Omega_n) = \vert\Omega_n\vert^{2m+1}\ ,
\label{eq:B.11}
\ee
and
\be
\chi''(\omega) = (-)^{m+1}\,\omega^{2m+1}\ ,
\label{eq:B.12}
\ee
We now need to distinguish between $m\geq 0$ and $m<0$. For $m\geq 0$ the spectrum is analytic,
the reactive part vanishes,
\be
\chi'(\omega) = 0\ ,
\label{eq:B.13}
\ee
and there is no long-time tail in the real-time domain. However, there is a long-time tail in the limit
of large imaginary time $\tau\to\infty$. $\chi(\tau)$ is given by
\be
\chi(\tau) = T\sum_{i\Omega_n} e^{-i\Omega_n\tau}\,\chi(i\Omega_n)\ .
\label{eq:B.14}
\ee
At $T=0$ the sum turns into an integral and we find
\be
\chi(\tau\to\infty) = \frac{1}{\pi}\,(-)^{m+1} (2m+1)!\,\frac{1}{\tau^{2(m+1)}}\ .
\label{eq:B.15}
\ee
% At small but nonzero temperature the asymptotic long-$\tau$ behavior is exponential, and the long-time tail crosses over to the true asymptotic behavior on a time scale given by
%\be
%\tau^* = \hbar/2\pi k_{\text B} T\ .
%\label{eq:B.15}
%\ee
% We note that in order for $\tau^*$ to be macroscopic one needs to go to extremely low temperatures.
An example for this behavior is the ferromagnet with damped magnons in $d=2$ with a conserved
order parameter, see Sec.~\ref{subsec:III.C}.

For $m<0$ the spectrum is singular at $\omega=0$ and there is a long-time tail even in the
real-time domain. We consider only $m=-1$, in
which case
\be
\chi'(\omega) = \delta(\omega)\ ,
\label{eq:B.16}
\ee
and the long-real-time behavior is a constant,
\be
\chi(t) = -i\ .
\label{eq:B.17}
\ee
An example is the antiferromagnet in $d=2$, see Sec.~\ref{subsubsec:II.B.1}.

%\bibliography{chi_L}

\begin{thebibliography}{54}
\bibitem[{2d_()}]{2d_footnote}
\bibinfo{note}{For $d<2$ there is no long-range order in a classical magnet,
  and hence there are no Goldstone modes.}
\bibitem[{not()}]{notation_footnote}
\bibinfo{note}{Throughout this paper, $a\propto b$ means ``$a$ is proportional
  to $b$", and $a\sim b$ means ``$a$ scales as $b$''.}
\bibitem[{FM_({\natexlab{a}})}]{FM_caveat}
\bibinfo{note}{This is true for a Heisenberg model of spins on a lattice. In
  itinerant ferromagnets there are many more fluctuating degrees of freedom and
  the situation is different. We will come back to this point in
  Secs.~\ref{sec:III} and \ref{sec:IV}.}
\bibitem[{dam()}]{damping_footnote}
\bibinfo{note}{There are some subtleties in this context that have to do with
  differences between insulating and itinerant magnets, see the discussion in
  Sec.~\ref{sec:IV}.}
\bibitem[{int()}]{integral_footnote}
\bibinfo{note}{The contribution is zero if one integrates over all frequencies,
  which requires an ultraviolet cutoff on the momentum integral. With an
  ultraviolet frequency cutoff it is an analytic function of the external
  frequency or wave number.}
\bibitem[{cla()}]{classical_limit_footnote}
\bibinfo{note}{Restoring $\hbar$, we see that this is also the form of the Bose
  distribution function in the classical limit, $\hbar\to 0$. Subtracting this
  classical contribution leads to Eq.~(\ref{eq:2.15}) with $n(x)$ replaced by
  $N(x) = n(x) - T/x$. The momentum integral is then infrared convergent in all
  dimensions. This is another way of saying that there is no singularity in a
  quantum ferromagnet.}
\end{thebibliography}

\end{document}